\documentclass[1p,11pt]{elsarticle}

\usepackage{pifont}
\usepackage{natbib}
\usepackage{geometry}
\usepackage{fleqn}
\usepackage{graphicx}

\usepackage{amssymb,amsmath}
\usepackage{color}
\usepackage{bm}

\DeclareMathAlphabet{\mathbi}{OML}{cmm}{b}{it} 
\newcommand{\non}{\nonumber}
\newtheorem{theorem}{Theorem}

\newcommand{\bx}{\mathbi{x}}

\newcommand{\bel}{\begin{equation}\label}
\newcommand{\ee}{\end{equation}}
\newcommand{\beq}{\begin{eqnarray}\label} 
\newcommand{\eeq}{\end{eqnarray}} 
\newcommand{\bc}{\begin{center}} 
\newcommand{\ec}{\end{center}} 
\newcommand{\ben}{\begin{enumerate}}
\newcommand{\een}{\end{enumerate}}
\newcommand{\bit}{\begin{itemize}}
\newcommand{\eit}{\end{itemize}}
\newcommand{\IV}{\int_{\mathcal{V}}}

\newcommand{\bdf}{\mathbi{f}}

\newcommand{\bu}{\mbox{\boldmath$u$}}

\newcommand{\bom}{\mbox{\boldmath$\omega$}}
\newcommand{\bphi}{\mbox{\boldmath$\phi$}}

\newcommand\shalf{\ensuremath{{\scriptstyle\frac{1}{2}}}}

\newcommand\quart{\ensuremath{{\scriptstyle\frac{1}{4}}}}

\newcommand\threehalves{\ensuremath{{\scriptstyle\frac{3}{2}}}}
\newcommand\fivehalves{\ensuremath{{\scriptstyle\frac{5}{2}}}}

\newcommand\onefifth{\ensuremath{{\scriptstyle\frac{1}{5}}}}

\newsavebox{\astrutbox}
\sbox{\astrutbox}{\rule[-5pt]{0pt}{25pt}}

\begin{document}

\title{\textbf{\small The role of BKM-type theorems in $3D$ Euler, Navier-Stokes 
and Cahn-Hilliard-Navier-Stokes analysis}}

\bc
{\em Dedicated to Prof. Edriss S. Titi on the on the occasion of his 60th birthday}
\ec

\author{John D. Gibbon\corref{cor1}}
\ead{j.d.gibbon@ic.ac.uk}
\address{Department of Mathematics, Imperial College London, London SW7 2AZ, UK.}

\author{Anupam Gupta}
\ead{anupam1509@gmail.com}
\address{FERMaT, Universit\'e de Toulouse, CNRS, INPT, INSA, UPS, Toulouse, France}

\author{Nairita Pal}
\ead{nairitap2009@gmail.com}
\address{Centre for Condensed Matter Theory, Department of Physics, Indian Institute of Science, Bangalore, 560 012, India.}

\author{Rahul Pandit}
\ead{rahulpandi@gmail.com}
\address{Centre for Condensed Matter Theory, Department of Physics, Indian Institute of Science, Bangalore, 560 012, India.}

\cortext[cor1]{Corresponding author}

\begin{abstract}
The Beale-Kato-Majda theorem contains a single criterion that controls the behaviour of solutions of the $3D$ 
incompressible Euler equations. Versions of this theorem are discussed in terms of the regularity issues surrounding 
the $3D$ incompressible Euler and Navier-Stokes equations together with a phase-field model for the statistical 
mechanics of binary mixtures called the $3D$ Cahn-Hilliard-Navier-Stokes (CHNS) equations. A theorem 
of BKM-type is established for the CHNS equations for the full parameter range. Moreover, for this latter set, it is 
shown that there exists a Reynolds number and a bound on the energy-dissipation rate that, remarkably, reproduces 
the $Re^{3/4}$ upper bound on the inverse Kolmogorov length normally associated with the Navier-Stokes equations 
alone. An alternative length-scale is introduced and discussed, together with a set of pseudo-spectral computations 
on a $128^{3}$ grid.
\end{abstract}


\date{\today} 

\maketitle


\section{Introduction}

\subsection{The $3D$ Euler, Navier-Stokes and Cahn-Hilliard-Navier-Stokes equations}

The fine-scale turbulent dynamics, commonly observed in numerical simulations and experiments, has long been thought to 
be related to the issues concerning the regularity of solutions of both the $3D$ incompressible Euler and Navier-Stokes equations, although these issues remain largely unresolved  
\cite{BardosTiti2007,MajdaBert2001,JDG2008,CFM1996,Leray1934,CF1988,FMRT2001,RRS2016,ESS2003,DGbook1995,Titi1987,Titi1990}. 
Respectively, these equations are 
\bel{euldef}
\left(\partial_{t}+\bu\cdot\nabla\right)\bu = - \nabla p\,,
\ee
and 
\bel{nsedef}
\left(\partial_{t}+\bu\cdot\nabla\right)\bu = \nu \Delta\bu - \nabla p + \bdf(\bx)\,.
\ee
In \eqref{euldef} and \eqref{nsedef} $\bu$ is a divergence-free
($\mbox{div}\,\bu =0$) velocity field, $\nu$ is the viscosity and $\bdf(\bx)$
is a divergence-free, mean-zero, $L^{\infty}$-bounded forcing. In this paper
the domain $\mathcal{V}$ is taken to be a periodic box of side $L$ and the 
uniform density $\rho$ is set to unity. 
\par\smallskip
Another system in which turbulent dynamics occurs is the phase-field model governed by the $3D$ Cahn-Hilliard 
equations. These are fundamental in the study of the statistical mechanics of binary mixtures 
\cite{CH1958,Chaikin2000,Hohenberg1977,Gunton1983,Bray1994,MuskatAM2012,MuskatARMA2013,Puri2009,Lothe1962,Onuki2002,Badalassi2003,
Perlekar2014,Cahn1961,Berti2005,Boffetta2012,Cabot2006,Celani2009,Pal2016,Gupta2014,Scarbolo2013,Yue2004}
\bel{chns1}
\partial_{t}\phi = \gamma\Delta \mu\,,
\ee
where the chemical potential $\mu = \delta\mathcal{F}/\delta\phi$ is related 
to the free energy 
\bel{Fdef}
\mathcal{F} = \IV \left[\frac{\Lambda}{2}|\nabla\phi|^{2} 
+ \frac{\Lambda}{4\xi^2}\left(\phi^{2}-1\right)^{2}\right]\,dV\,.
\ee
$\mu$ is thus given by 
\bel{chns2}
\mu = \Lambda\left[-\Delta\phi + \xi^{-2}\left(\phi^{3} - \phi\right)\right]\,.
\ee
This model can be used to study the mixing of two fluids, which are \textit{immiscible} below a 
critical temperature, via a phase field $\phi$. In equilibrium, $\phi = -1$ for one phase and 
$\phi = 1$ for the other.  The advantage of such a model is the continuity of the thin interface, 
of thickness $\xi$, between the two fluids. The existence of this interface removes the 
necessity of dealing with the complications of tracking a free boundary. When \eqref{chns1}
is coupled to the $3D$ Navier-Stokes equations ($\mbox{div}\,\bu =0$)
\beq{chns3}
\left(\partial_{t} + \bu\cdot\nabla\right)\phi &=& \gamma\Delta \mu\,,\qquad\qquad \mbox{div}\,\bu = 0\\
\left(\partial_{t} + \bu\cdot\nabla\right)\bu &=& \nu \Delta \bu - \phi \nabla\mu - \nabla p + \bdf(\bx)\,,\label{chns3A}
\eeq
the combination of \eqref{chns2}, \eqref{chns3} and \eqref{chns3A} are known as the Cahn-Hilliard-Navier-Stokes (CHNS) 
equations. The parameter $\gamma$ in \eqref{chns1} is called the mobility (Bray \cite{Bray1994}), and $\xi$ is the interface 
thickness. The interfacial dynamics are of especial interest, particularly regarding the immiscible Rayleigh-Taylor 
instability (RTI), which is manifest in this thin mixing layer\,: for references on the ubiquity of the RTI see  
\cite{Petrasso1994,Taleyarkhan2002,Munk1998,Bhatnagar2014,Waddell2001,Ramaprabhu2006,Dalziel2008,Rao2016,Livescu2013}.
Whether tightly-packed interfacial level sets remain continuous as time evolves is a question that is closely connected to 
the issue of the regularity of solutions, which remains an open problem for all these three sets of equations in three dimensions ($3D$). Various results are known in two dimensions ($2D$), such as the regularity of not only the $2D$ 
Navier-Stokes equations \cite{CF1988,FMRT2001,DGbook1995,Titi1987,Titi1990} but also of the stand-alone $2D$ Cahn-Hilliard equations (Elliott and Songmu \cite{Elliott1986}). The regularity problem for the $2D$ Cahn-Hilliard-Navier-Stokes (CHNS) equations has been solved in some remarkable papers by Abels \cite{Abels2009a,Abels2009b} and Gal and Grasselli \cite{Gal2010} using different boundary conditions. In $3D$, however, the issue remains a formidable open problem. Nevertheless, in the light of criteria that control their regularity, they do possess certain features in common with both the Euler and Navier-Stokes equations, and it is these that are the subject of this paper. 


\subsection{Statement of a theorem of BKM-type for the CHNS equations}

The fundamental theorem that governs the behaviour of solutions of the $3D$ Euler equations is called the 
Beale-Kato-Majda (BKM) theorem \cite{BKM1984}\,: see also Bardos and Titi \cite{BardosTiti2007} and Gibbon 
\cite{JDG2008}. The statement of the theorem is simple. For $n \geq 0$, let us define
\bel{Hndef}
H_{n} = \IV |\nabla^{n}\bu|^{2}\,dV\,.
\ee
Now consider the vorticity $\bom = \mbox{curl}\,\bu$. The notation $\|\cdot\|_{p} = \left(\IV |\cdot|^{p}dV\right)^{1/p}$ 
means that $\|\bom\|_{\infty}$ is the maximum or sup-norm of the vorticity in the domain $\mathcal{V}$. 
\begin{theorem}\label{BKMthm} (Beale, Kato and Majda \cite{BKM1984}) 
For initial data of the $3D$ Euler equations satisfying $u_{0} \in H_{n}$ for $n \geq 3$, suppose there exists a 
solution on the interval $[0,\,T^{*})$ that loses regularity at the earliest time $T^{*}$, then 
\bel{BKM1}
\int_{0}^{T^*}\|\bom\|_{\infty}\,d\tau = \infty\,.
\ee
Conversely, if, for every $T>0$, $\int_{0}^{T}\|\bom\|_{\infty}\,d\tau < \infty$, then solutions of the $3D$ Euler 
equations remain regular on $[0,\,T]$. 
\end{theorem}
\par\medskip\noindent
The proof in \cite{BKM1984} is short and the strategy is by contradiction. After some work, BKM found a differential 
inequality for $H_n$, in terms of $\|\bom\|_{\infty}$ which, when integrated in time up to and including $T^*$, 
proves that $H_n$ is controlled from above by $\int_{0}^{T^*}\|\bom\|_{\infty}\,d\tau$. Given that the theorem 
presupposes that $H_n$ loses regularity at $T^*$, we cannot have $H_{n}(T^{*})=\infty$ while $\int_{0}^{T^*}
\|\bom\|_{\infty}\,d\tau$ remains finite.  
\par\smallskip
Compared to the $3D$ Navier-Stokes equations, little is known about the behaviour\footnote{The wild solutions 
of De Lellis and Szekelyhidi \cite{DeLSz2009,DeLSz2013} lie in a category of their own\,: see also Buckmaster and Nicol 
\cite{BV2017}.} of solutions of the $3D$ Euler equations \cite{BardosTiti2007}. The value of the BKM theorem is that it furnishes us with a single, numerically testable criterion based on the behaviour of the time integral $\int_{0}^{T}\|\bom\|_{\infty}\,d\tau$. There is a long history of numerical experiments that have aimed to test whether a singularity develops (see the list in \cite{JDG2008}) but the latest work suggests that solutions do not blow up but undergo double exponential growth \cite{DHY2005,Kerr2013}. The theorem also rules out potential 
algebraic singularities of a certain type\,: for instance, if one performs a numerical simulation and observes a singularity 
of the type $\|\bom\|_{\infty} \sim (T^* - t)^{-p}$, then $\int_{0}^{T^*}\|\bom\|_{\infty}\,d\tau$ is finite for $0 <p<1$. 
The theorem says that no singularity can occur, whereas the claim is that one has been observed.  The ensuing contradiction 
can only be resolved by realizing that the observed singularity is an artefact of the numerical scheme employed. True singularities of this type must have $p\geq 1$.
\par\smallskip
Theorem \ref{BKMthm} is specific to the $3D$ Euler equations and centres around the $\|\bom\|_{\infty}$-criterion in 
(\ref{BKM1}), but it is possible to widen this idea to other model problems which display similar criteria for loss 
of regularity. These we will label as being of ``BKM-type''. In fact, two theorems of BKM-type that already been proved. 
The first is for the stochastic Euler equations by Crisan, Flandoli and Holm \cite{CFH2017}. The second is a theorem 
similar to Theorem \ref{BKMthm} that has already been proved by the authors in \cite{Gibbon2017PRE} for the $3D$-CHNS 
equations, but with unit parameters only. One of the aims of this paper, among others, is to extend this proof to the 
full parameter range and to discuss its relationship with the versions valid for the $3D$ Euler and Navier-Stokes 
equations. Before stating it here, some background is necessary. The energy of the full CHNS system is given by (see 
Celani \textit{et al.} \cite{Celani2009})
\bel{Edefintro}
E(t) = \IV \left\{\shalf\Lambda|\nabla\phi|^{2} + \frac{\Lambda}{4\xi^{2}} (\phi^{2} - 1)^{2} 
+ \shalf|\bu|^{2}\right\}\,dV\,.
\ee
This is comprised of a sum of $L^{2}$-norms and clearly suggests an $L^\infty$-equivalent denoted as $E_{\infty}$ 
and defined by
\bel{Einfintro}
E_{\infty}(t) = \shalf\Lambda\|\nabla\phi\|_{\infty}^{2} + \frac{\Lambda}{4\xi^{2}} (\|\phi\|_{\infty}^{2} - 1)^{2} 
+ \shalf\|\bu\|_{\infty}^{2}\,.
\ee
We also need a similar definition similar to $H_{n}$ involving $\phi$ 
\bel{Pndef}
P_{n} = \IV |\nabla^{n}\phi|^{2}\,dV\,.
\ee
The statement of the theorem for the full parameter range follows here below and its proof is discussed in 
\S\ref{CHNSBKM} and \ref{appB}\,:
\begin{theorem}\label{thmCHNS}
Consider the $3D$ CHNS equations on a periodic domain $\mathcal{V} = [0,\,L]^{3}$. For initial data $u_{0} \in H_{n}$, 
for $n \geq 2$, and $\phi_{0}\in P_{n}$, for $n \geq 3$, suppose there exists a solution on the interval $[0,\,T^{*})$, 
where $T^{*}$ is the earliest time that the solution loses regularity, then
\bel{thm1a}
\int_{0}^{T^{*}}E_{\infty}(\tau)\,d\tau = \infty\,.
\ee
Conversely, there exists a global solution of the 3D CHNS equation if, for every $T > 0$,
\bel{thm1b}
\int_{0}^{T}E_{\infty}(\tau)\,d\tau < \infty\,.
\ee
\end{theorem}
\par\smallskip\noindent
Clearly, this theorem is of BKM-type where $E_{\infty}$ replaces $\|\bom\|_{\infty}$ in Theorem \ref{BKMthm}. As 
in the BKM theorem above, it provides us with a precise, single criterion for numerically monitoring the blow-up 
of solutions. Some types of blow-up could potentially be extremely subtle, such as a cusp forming in a tightly 
packed level sets in the CHNS-interface\,; this could potentially cause a high derivative to become singular. 
These are ruled out if $\int_{0}^{t}E_{\infty}\,d\tau <\infty$.  However, an obvious question to ask is why the 
Navier-Stokes part of $E_{\infty}$ is proportional to $\|\bu\|_{\infty}^{2}$ and not $\|\bom\|_{\infty}$? This 
question is answered in \S\ref{NSsum} where several well-known $3D$ Navier-Stokes regularity criteria are summarized 
(see Table 1) and where it is shown that while $E_{\infty}$-theorem is akin to the Euler equations in being of 
BKM-type, the $\|\bu\|_{\infty}^{2}$ term has its origins in the Navier-Stokes equations. This is followed by a 
section on the CHNS equations, in which some new results on bounds for the energy dissipation rate in terms of 
the Reynolds number are displayed. 
\par\smallskip
In the original proof of the $E_{\infty}$-theorem in \cite{Gibbon2017PRE}, the parameters $\nu,~\Lambda,~\gamma$ 
and $\xi$ were set to unity for convenience. In \S\ref{CHNS}, dimensional analysis is used to create a new version 
of the proof with the full parameter range.
\par\smallskip
Finally, thanks to our state-of-the-art direct numerical simulations (DNSs) in \S\ref{altls}, we have been able to 
monitor the complete time series of the energy dissipation rate and thus calculate the mean dissipation rate. These 
DNSs also help us to estimate a new alternative length scale based on $\sqrt \Lambda$. This helps us to see if there 
is (or is not) any ordering of the conventional length scale and the new alternative scale\,; this cannot be predicted analytically.


\section{Regularity properties of the $3D$ Navier-Stokes equations\,:\\
the $\int_{0}^{t}\|\bu\|_{\infty}^{2}\,d\tau$ criterion}\label{NSsum}

The structure of $E_{\infty}$ in Theorem 2 is intriguing and raises the question why the Navier-Stokes 
contribution is of the form $\|\bu\|_{\infty}^{2}$ and not the conventional $\|\bom\|_{\infty}$. The 
first subsection discusses this question while the second summarizes current knowledge of the boundedness 
of time-averages, particularly the energy dissipation rate which is of relevance when this issue is raised 
for the $3D$ CHNS equations in \S\ref{CHNSenergy}. 

\subsection{$\int_{0}^{t}\|\bu\|_{\infty}^{2}\,d\tau$ as a Navier-Stokes regularity criterion}
\medskip
$3D$ Navier-Stokes and Euler regularity are substantially different in that pointwise control in time over 
$H_{1}$ is sufficient for the existence and uniqueness of solutions of the $3D$ Navier-Stokes equations 
whereas this is insufficient for the $3D$ Euler equations which require the finiteness of $\|\bom\|_{\infty}$. 
To look further at this, let us formally differentiate\footnote{See \cite{FMRT2001,RRS2016} for a more rigorous 
weak solution approach.} $H_{1}$ with respect to time to obtain\,:
\bel{H1a}
\shalf \dot{H}_{1} \leq - \nu H_{2} + \left|\IV \bom\cdot(\bom\cdot\nabla\bu)\,dV\right| + \|\bdf\|_{2}H_{1}^{1/2}\,.
\ee
There are two ways of estimating the central integral term\,:
\bel{H1aex}
\left|\IV \bom\cdot(\bom\cdot\nabla\bu)\,dV\right| \leq 
\left\{
\begin{array}{l}
\|\bom\|_{\infty}H_{1}\,,\\
\|\bu\|_{\infty}H_{1}^{1/2}H_{2}^{1/2}\,.
\end{array}
\right.
\ee
With the first estimate, (\ref{H1a}) becomes
\bel{H1b}
\shalf \dot{H}_{1} \leq - \nu H_{2} + \|\bom\|_{\infty}H_{1} + \|\bdf\|_{2}H_{1}^{1/2}\,,
\ee
and with the second, 
\bel{H1c}
\shalf \dot{H}_{1} \leq - \shalf\nu H_{2} + \shalf\nu^{-1}\|\bu\|_{\infty}^{2}H_{1} + \|\bdf\|_{2}H_{1}^{1/2}\,.
\ee
Dropping the negative $H_{2}$-terms in both (\ref{H1b}) and (\ref{H1c}) it is clear that $H_{1}(t)$ is bounded 
from above provided either 
\bel{H1f}
\int_{0}^{t} \|\bom\|_{\infty}\,d\tau < \infty  \qquad\mbox{or}\qquad \int_{0}^{t}\|\bu\|_{\infty}^{2}\,d\tau < \infty\,.
\ee
The first is obviously the BKM criterion of Theorem \ref{BKMthm}, valid for both the $3D$ Euler and Navier-Stokes equations, 
but the second is valid only for $3D$ Navier-Stokes because of the role played by the viscous term in deriving (\ref{H1c}). 
It is the second criterion that appears naturally in $E_{\infty}$, as the proof in \ref{appB} shows.

\begin{table}\label{tab1}
\bc
\begin{tabular}{||c|c||}\hline
What is known                                   &  What is sufficient for regularity \\\hline
$\|\bu(\cdot,\,t)\|_{2} < \infty$ & $\|\bu(\cdot,\,t)\|_{3} < \infty$\\\hline
$\int_{0}^{t}\|\bu\|_{\infty}\,d\tau < \infty$ &  $\int_{0}^{t}\|\bu\|^{2}_{\infty}\,d\tau < \infty$\\\hline
$\int_{0}^{t}H_{1}\,d\tau < \infty$  & $\int_{0}^{t}H_{1}^{2}\,d\tau < \infty$ \\\hline
$\int_{0}^{t}\|\bom\|^{1/2}_{\infty}\,d\tau < \infty$ &  $\int_{0}^{t}\|\bom\|_{\infty}\,d\tau < \infty$\\\hline
\end{tabular}\label{table1}
\ec
\caption{\scriptsize Table of the results that are known (left column) for the $3D$ Navier-Stokes equations and 
those results that are sufficient for regularity but unproved (right column). The notation is $\|\cdot\|_{p} = 
\left(\IV |\cdot|^{p}\,dV\right)^{1/p}$. The results $\int_{0}^{t}\|\bu\|_{\infty}\,d\tau < \infty$ and 
$\int_{0}^{t}\|\bom\|^{1/2}_{\infty}\,d\tau < \infty$ are both due to Guillop\'e, Foias, and Temam \cite{FGT1981}.} 
\end{table}
\par\smallskip
The alternative criterion, $\int_{0}^{t}\|\bu\|_{\infty}^{2}\,d\tau < \infty$, displayed in (\ref{H1f}) has a place 
in the broader class of regularity criteria due to Serrin (see \cite{RRS2016})
\bel{Serrin}
u \in L^{p}\left(0,T\,;\, L^{q}\right)\,,\qquad\qquad 2/p + 3/q = 1\,.
\ee
Table 1 displays a set of Navier-Stokes regularity criteria, the first row of which contains the $\|\bu(\cdot,\,t)\|_{3} 
< \infty$ criterion of Escauriaza, Seregin and Sverak \cite{ESS2003}\,: it is clear that this criterion lies at one end 
of (\ref{Serrin}) with $p=\infty$ and $q=3$, while the $\int_{0}^{t}\|\bu\|^{2}_{\infty}\,d\tau < \infty$ criterion, 
which lies in the second row, lies at the other where $p=2$ and $q=\infty$. Finally we note that there are other Navier-Stokes 
regularity criteria that lie outside this class\,: for instance, those based on the pressure \cite{Tran2016}.


\subsection{Bounded time averages}
\medskip
The Navier-Stokes equations possess a well known energy inequality that goes back to Leray \cite{Leray1934}. 
It takes the form\footnote{The mathematical statements in this section are purely formal\,: for a full weak 
solution exposition based on Leray's weak solutions \cite{Leray1934}, see \cite{CF1988,FMRT2001,RRS2016}.}
\bel{Ler1} 
\shalf \frac{d~}{dt}\IV|\bu|^{2}dV \leq -\nu H_{1} + \|\bu\|_{2}\|\bdf\|_{2}\,.
\ee
The $\bu\cdot(\bu\cdot\nabla\bu)$ term vanishes under the Divergence Theorem. With a time average defined by
\bel{tadef}
\left<\cdot \right>_{T} = \frac{1}{T}\int_{0}^{T}\cdot\,d\tau 
\ee
and with an average velocity $U$ and a box frequency $\varpi_{0}$ defined by 
\bel{Udef}
U^{2} = L^{-3}\left<\|\bu\|_{2}^{2}\right>_{T}\,,\qquad\qquad \varpi_{0} = \nu L^{-2}\,,
\ee
and with Grashof and Reynolds numbers defined as
\bel{GrRedef}
Gr = \frac{L^{3/2}\|\bdf\|_{2}}{\nu^2}\,,\qquad\qquad Re= \frac{UL}{\nu}\,,
\ee
a time average of \eqref{Ler1} gives 
\bel{Ler2}
\left<H_{1}\right>_{T} \leq \varpi_{0}^{2}L^{3} Gr Re + O\left(T^{-1}\right)\,.
\ee
Doering and Foias \cite{DF2002} have shown that, for a forcing function with a single scale $\ell$, for which we 
take $\ell= L$ for convenience, then $Gr \leq c\,Re^{2}$, where the dimensionless constant $c$ is a function of 
the shape of the forcing. Thus \eqref{Ler2} becomes 
\bel{Ler3}
\left<H_{1}\right>_{T} \leq c\,\varpi_{0}^{2}L^{3} Re^{3} + O\left(T^{-1}\right)\,,
\ee
and the energy-dissipation rate $\mathcal{E}$ is bounded by
\bel{endiss}
\mathcal{E} = \nu L^{-3} \left<H_{1}\right>_{T}\,,
\ee
and so we end up with the classic estimate for the inverse Kolmogorov length $\lambda_{k}^{-1}$ 
\bel{kol1}
L\lambda_{k}^{-1} = \left(\mathcal{E}/\nu^{3}\right)^{1/4}\qquad\Rightarrow\qquad
L\lambda_{k}^{-1} \leq c\,Re^{3/4}\,.
\ee
This type of $Re^{3/4}$-estimate is reflected in similar results for the $3D$ CHNS equations displayed below.

\section{The 3D CHNS equations}\label{CHNS}

\subsection{New estimates on the energy dissipation rate}\label{CHNSenergy}

Consider the energy of the full CHNS equations as in Celani, \textit{et al.}~\cite{Celani2009} 
stated earlier in (\ref{Edefintro})
\bel{Edef}
E(t) = \IV \left\{\shalf\Lambda|\nabla\phi|^{2} + \frac{\Lambda}{4\xi^{2}} (\phi^{2} - 1)^{2} + \shalf|\bu|^{2}\right\}\,dV\,.
\ee
Then a formal differentiation gives 
\beq{Ecal1}
\frac{dE}{dt} &=& \IV\left\{\Lambda\nabla\phi\cdot\nabla\left[-\bu\cdot\nabla\phi + \gamma\Delta \mu\right] + 
\left(\mu + \Lambda\Delta\phi\right)\left(-\bu\cdot\nabla\phi + \gamma\Delta\mu\right) \right\}\,dV\nonumber\\
&+& \IV \left\{\bu\cdot\left(\nu\Delta \bu - \phi \nabla\mu - \nabla p + \bdf \right)\right\}\,dV\,.
\eeq
Integrating by parts on the first term and using the property $\mbox{div}\,\bu = 0$ and the Divergence Theorem 
to remove the pressure term, we find
\beq{Ecal2}
\frac{dE}{dt} &=& - \Lambda\IV \Delta\phi\left[-\bu\cdot\nabla\phi + \gamma\Delta \mu\right]\,dV\nonumber\\
&+&  \IV\left(\mu + \Lambda\Delta\phi\right)\left(-\bu\cdot\nabla\phi + \gamma\Delta\mu\right)\,dV\nonumber\\
&-& \IV \phi\,\bu\cdot\nabla\mu\,dV +\IV\left(-\nu|\nabla\bu|^{2} + \bu\cdot\bdf\right)\,dV\,.
\eeq
Terms in the first and second line of (\ref{Ecal2}) cancel. Moreover, $\IV \bu\cdot\nabla(\phi\mu)\,dV = 0$ because 
$\mbox{div}\,\bu = 0$, leaving us with
\bel{Ecal3}
\frac{dE}{dt} = -\IV\left(\nu|\nabla\bu|^{2} + \gamma|\nabla\mu|^{2} \right)\,dV + \IV\bu\cdot\bdf\,dV\,.
\ee
Thus,  without any additive forcing, $dE/dT < 0$, in which case $E$ decays, a result which is true in every dimension. 
For the forced case we can do better with time averages 
$\left<\cdot\right>_{T}$ up to time $T$ defined in (\ref{tadef}).  Define the full energy-dissipation rate as
\bel{ta1}
\mathcal{E}(\nu,\gamma) =  L^{-3}\left<\IV\left(\nu|\nabla\bu|^{2} + \gamma|\nabla\mu|^{2} \right)\,dV\right>_{T} \,,
\ee
then a time average of (\ref{Ecal3}) gives
\bel{ta2}
\mathcal{E}(\nu,\gamma) \leq U\|\bdf\|_{2} + O\left(T^{-1}\right)
= \nu^{3}L^{-4} GrRe + O\left(T^{-1}\right) ,
\ee
where the average velocity $U$ is defined in (\ref{Udef}). Defining a Kolmogorov-like length in the conventional way 
\bel{Kol1}
L\lambda_{\nu,\gamma}^{-1} = \left(\mathcal{E}(\nu,\gamma) /\nu^{3}\right)^{1/4}\,,
\ee
we can appeal to the modified Doering-Foias relation between $Gr$ and $Re$ proved in the Appendix and shown earlier 
in (\ref{endiss}). Remarkably, this still stands for the CHNS system with minor modifications. Thus we have 
\bel{ta3}
L\lambda_{\nu,\gamma}^{-1} \leq c\,Re^{3/4} \,.
\ee
\begin{table}
\bc
\begin{tabular}{||c|c||}\hline\hline
Parameter & Dimension\\\hline\hline
$[\nu]$ & $L^{2}T^{-1}$\\\hline
$[\xi]$ & $L$\\\hline
$[\Lambda]$ & $L^{4}T^{-2}$\\\hline
$[\gamma]$ & $T$\\\hline
$[\phi]$ & none\\\hline
$[\mu]$ & $L^{2}T^{-2}$\\\hline\hline
\end{tabular}\label{table2}
\ec
\caption{Dimensions of the various parameters and the variable $\mu$ in \eqref{chns1} -- \eqref{chns2}.}
\end{table}


\subsection{An alternative length scale}\label{altls}

Because $\sqrt{\Lambda}$ and $\nu$ have the same dimensions it is also possible 
to define new variables $\mathcal{E}_{\tiny\Lambda}$, $L\lambda_{\Lambda,\gamma}^{-1}$ 
in which $\nu$ has been replaced by $\sqrt{\Lambda}$. Thus we define
\bel{Elamdef1}
\mathcal{E}(\Lambda,\gamma) =  L^{-3}\left<\IV\left(\sqrt{\Lambda}|\nabla\bu|^{2} 
+ \gamma|\nabla\mu|^{2} \right)\,dV\right>_{T} \,,
\ee
with 
\bel{Elamdef2}
L\lambda_{\Lambda,\gamma}^{-1} = \left(\mathcal{E}(\Lambda,\gamma) /\Lambda^{3/2}\right)^{1/4}\,.
\ee
Which is the smallest scale, $\lambda_{\nu,\gamma}$ or $\lambda_{\Lambda,\gamma}$? The answer depends on 
how large, in relative terms, are the integrals in (\ref{ta1}) and (\ref{Elamdef1}), and how large, in 
relative terms, are $\nu$ and $\sqrt{\Lambda}$?
\par\smallskip
To investigate these two scales, we turn to direct numerical simulations (DNSs) of the 3D CHNS equations, using constant energy injection and a constant forcing wave number. We use a pseudospectral method, with $N/2$ dealiasing because of the cubic nonlinearity and, for the purpose of illustration, $128^{3}$ collocation points. We use a second-order Adams-Bashforth method for time marching. In our CHNS description, both components of the fluids have the same viscosity; and we assume that $\gamma$ is independent of $\phi$. Other details of such a DNS can be found in Refs.\cite{Pal2016,Gibbon2017PRE} which we follow here.
\par
\begin{figure}[h] 
\includegraphics[width=0.32\linewidth]{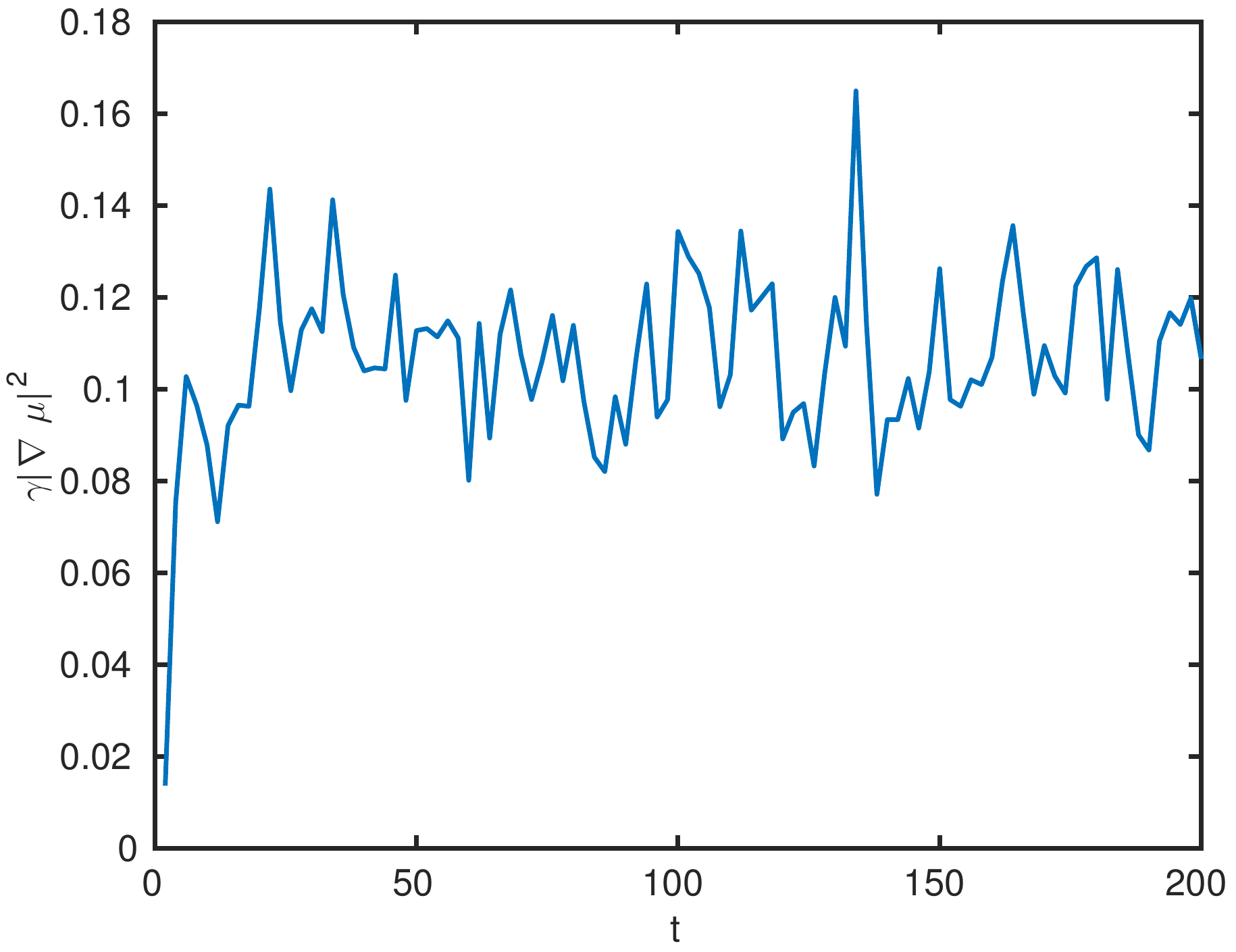}
\includegraphics[width=0.32\linewidth]{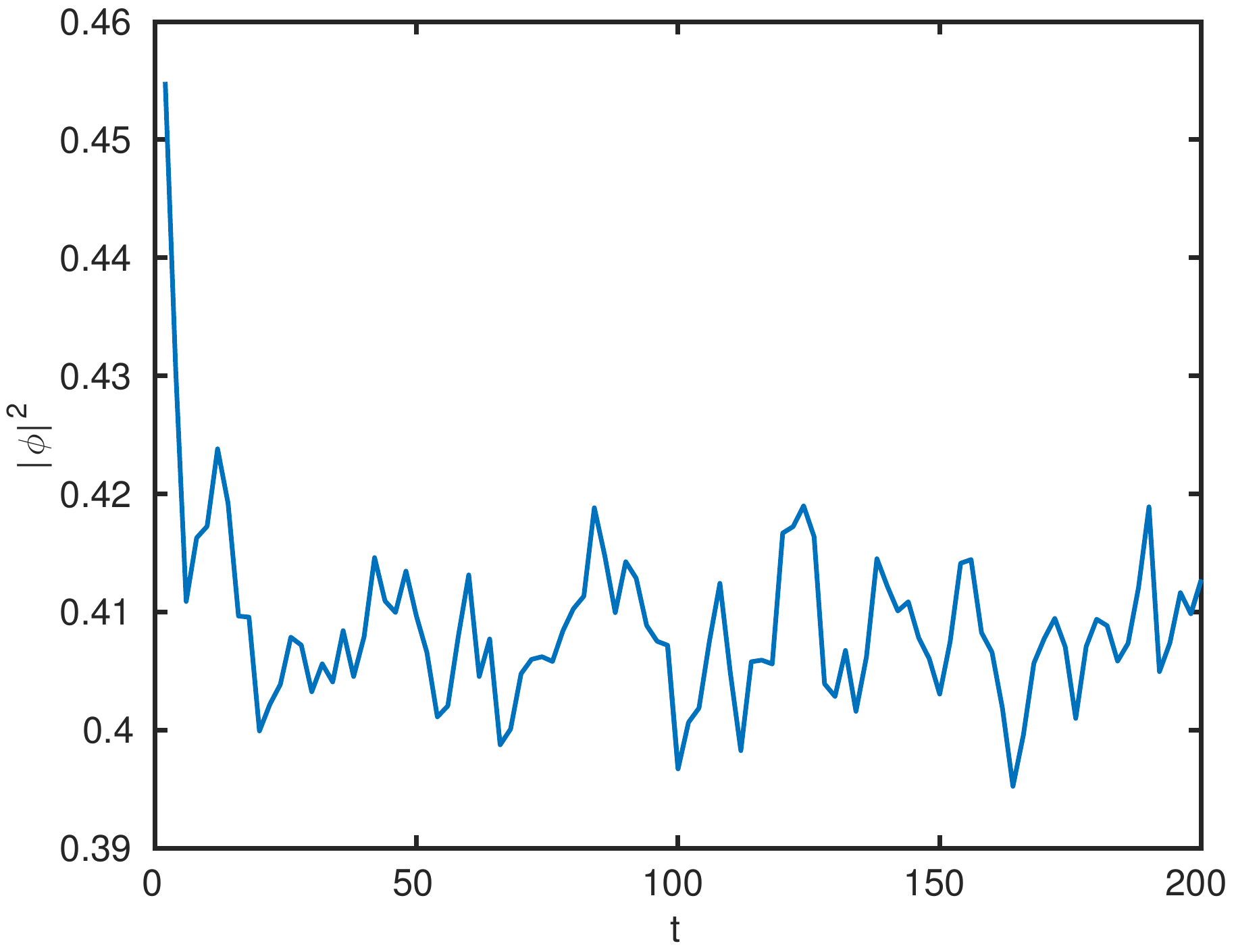}
\includegraphics[width=0.32\linewidth]{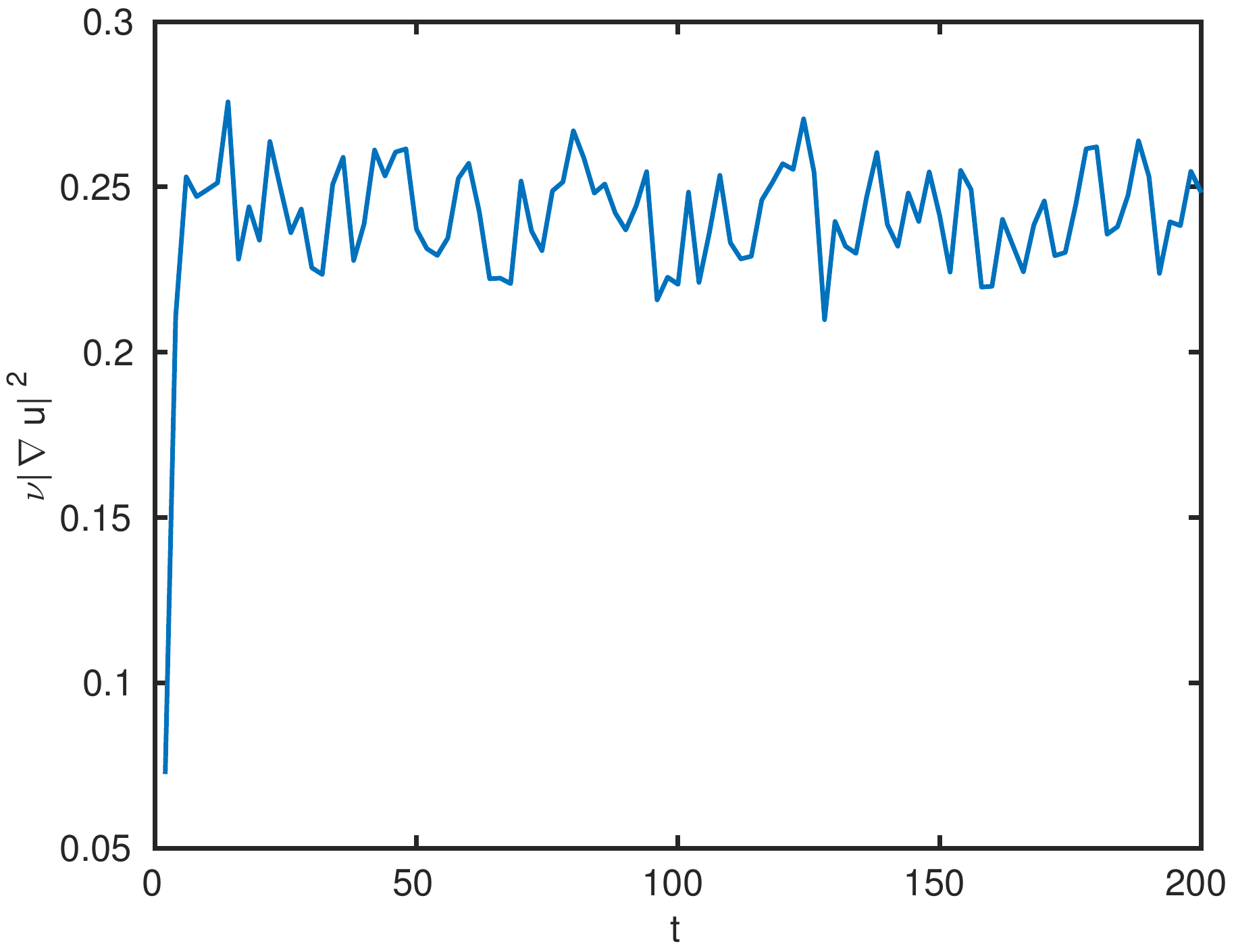}
\caption{Plots versus time $t$ of $\gamma\|\nabla \mu\|_{2}^2$ (left panel), $\|\phi\|_{2}^2$ (middle panel), and $\nu\|\nabla\bu\|_{2}^2$ (right panel) for run ${\tt R7}$. The labelling in the figures is 
$|\cdot|^{2}\equiv \|\cdot\|_{2}^{2}$.}\label{fig:R1}
\end{figure}

\begin{figure}[h]
\includegraphics[width=0.32\linewidth]{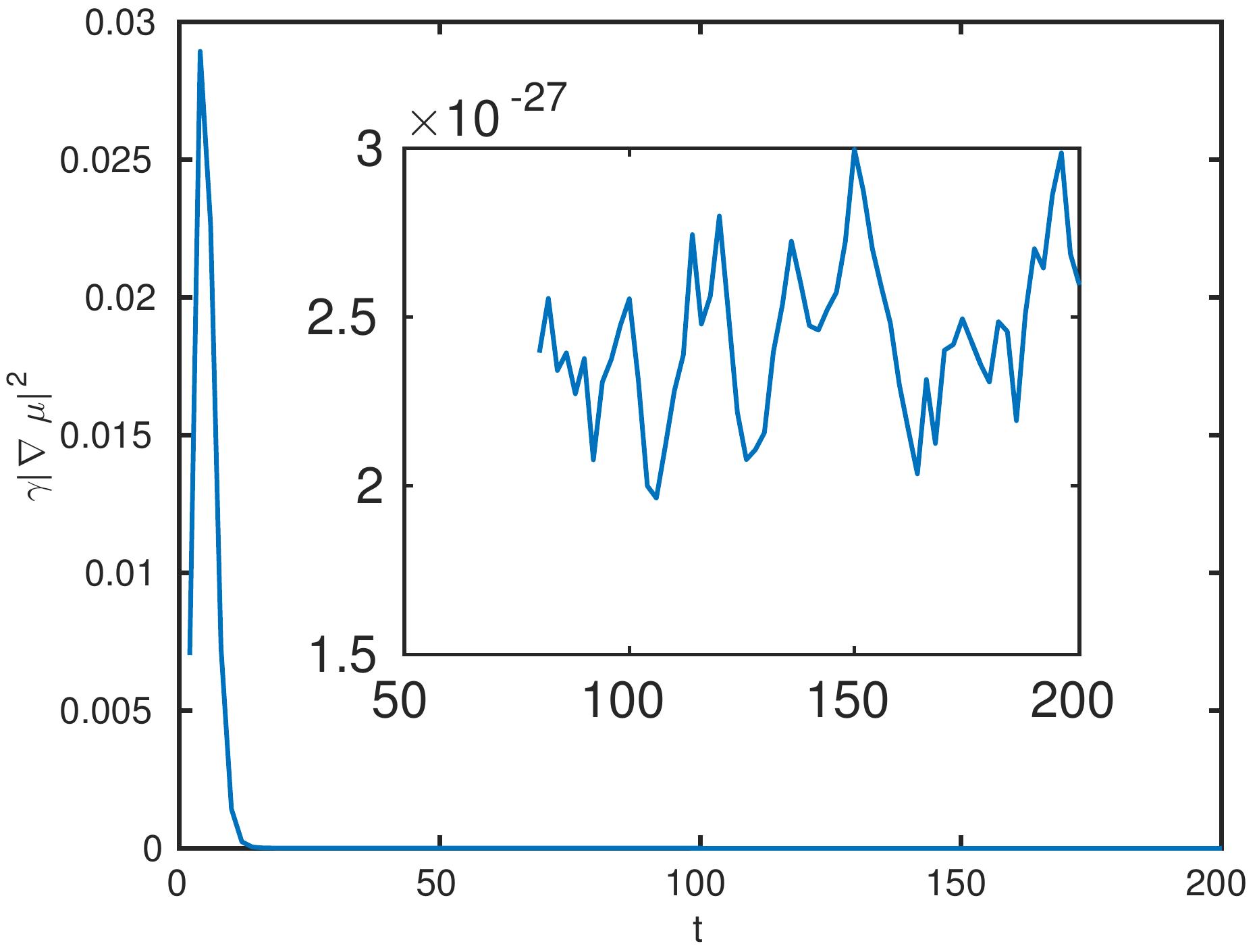}
\includegraphics[width=0.32\linewidth]{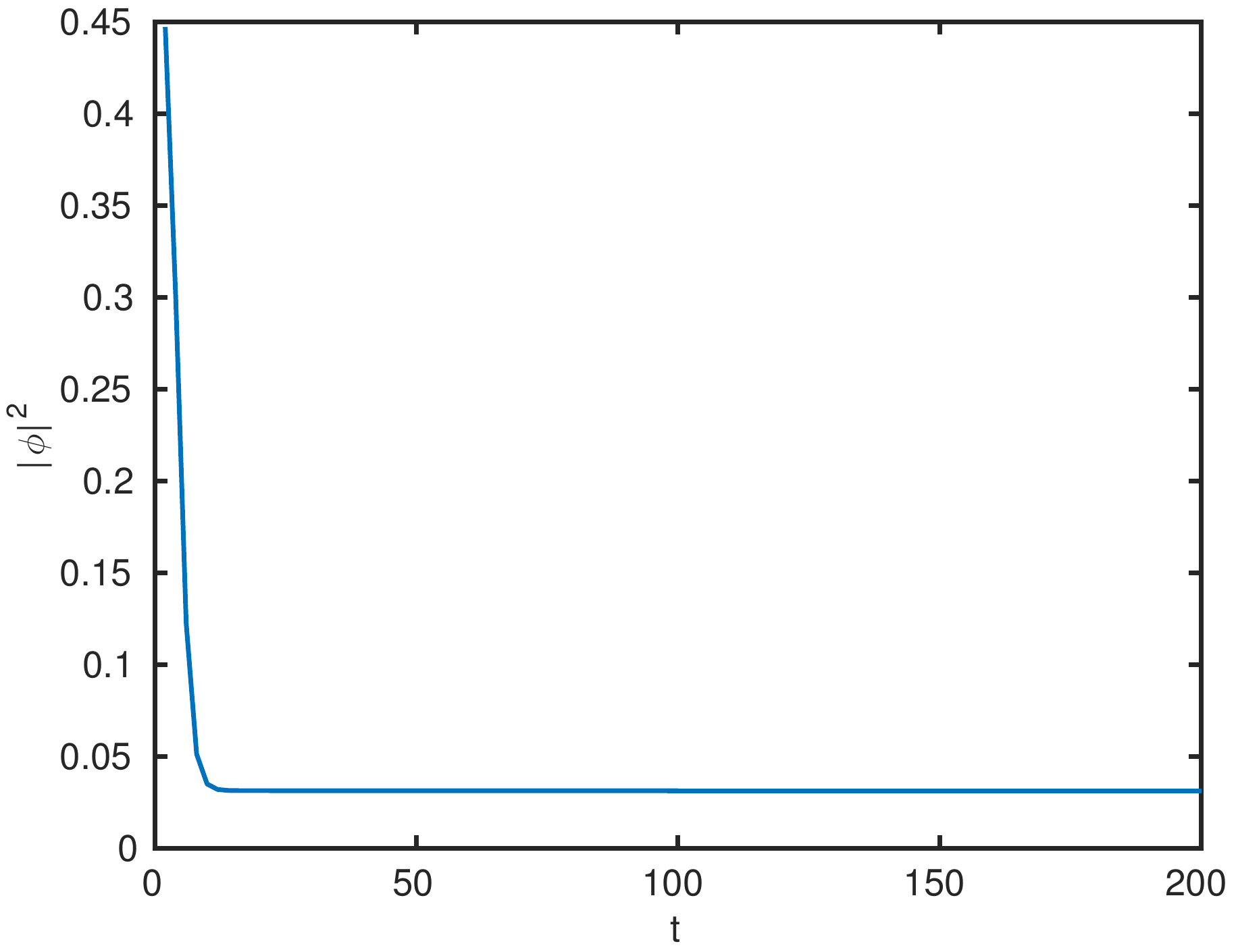}
\includegraphics[width=0.32\linewidth]{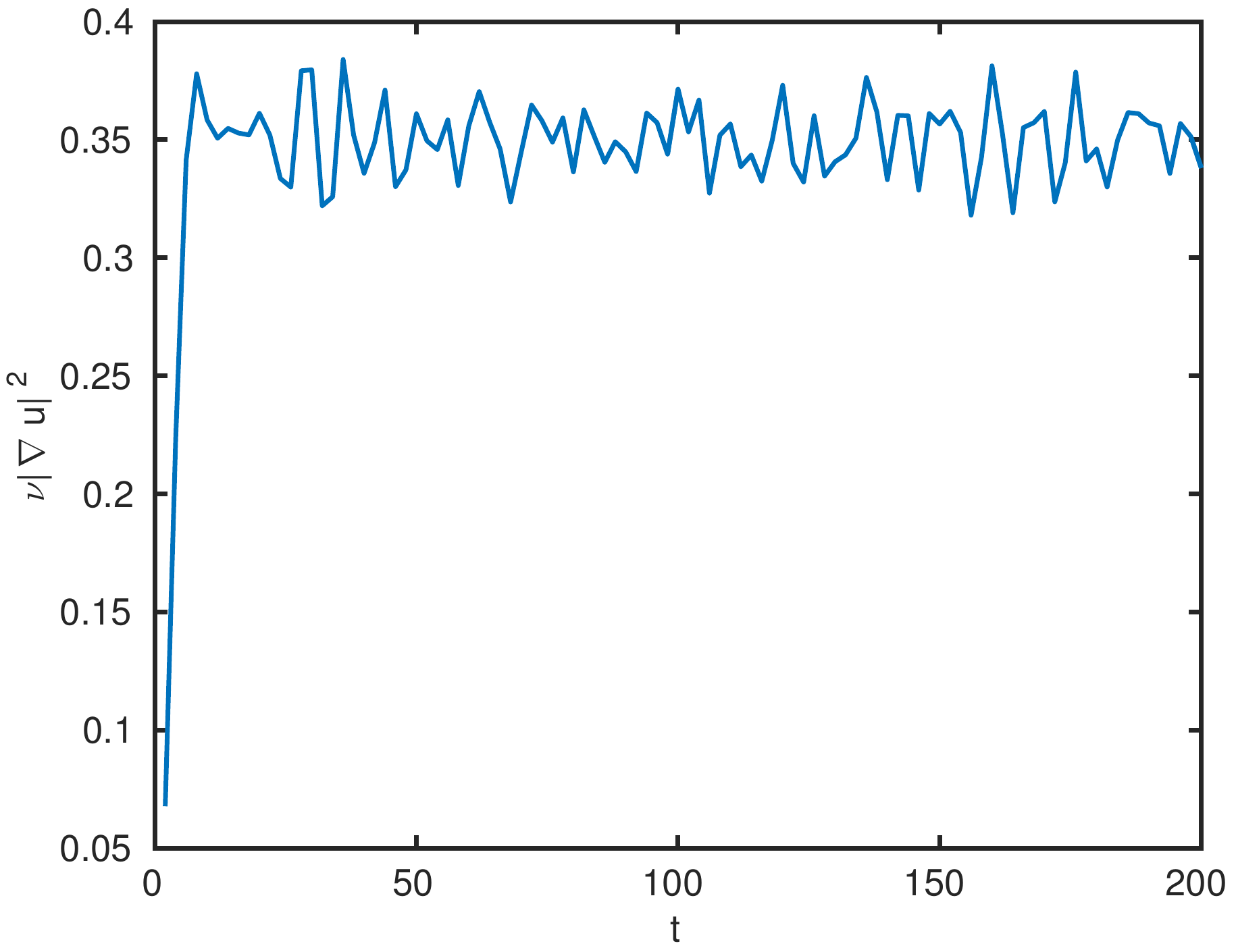}
\caption{Plots versus time $t$ of $\gamma\|\nabla \mu\|_{2}^2$ (left panel), $\|\phi\|_{2}^2$ (middle panel), 
and $\nu\|\nabla\bu\|_{2}^2$ (right panel) for run ${\tt R3}$. The inset in the left panel shows the plot of 
$\gamma\|\nabla\mu\|_{2}^2$ versus time $t$\,; this shows that $\gamma\|\nabla\bu\|_{2}^2$ is vanishingly small 
for this run, in the statistically steady state. As in Fig. \ref{fig:R1}, the labelling in the figures is 
$|\cdot|^{2}\equiv \|\cdot\|_{2}^{2}$.}\label{fig:R3}
\end{figure}

\begin{figure}[h]
\includegraphics[width=0.32\linewidth]{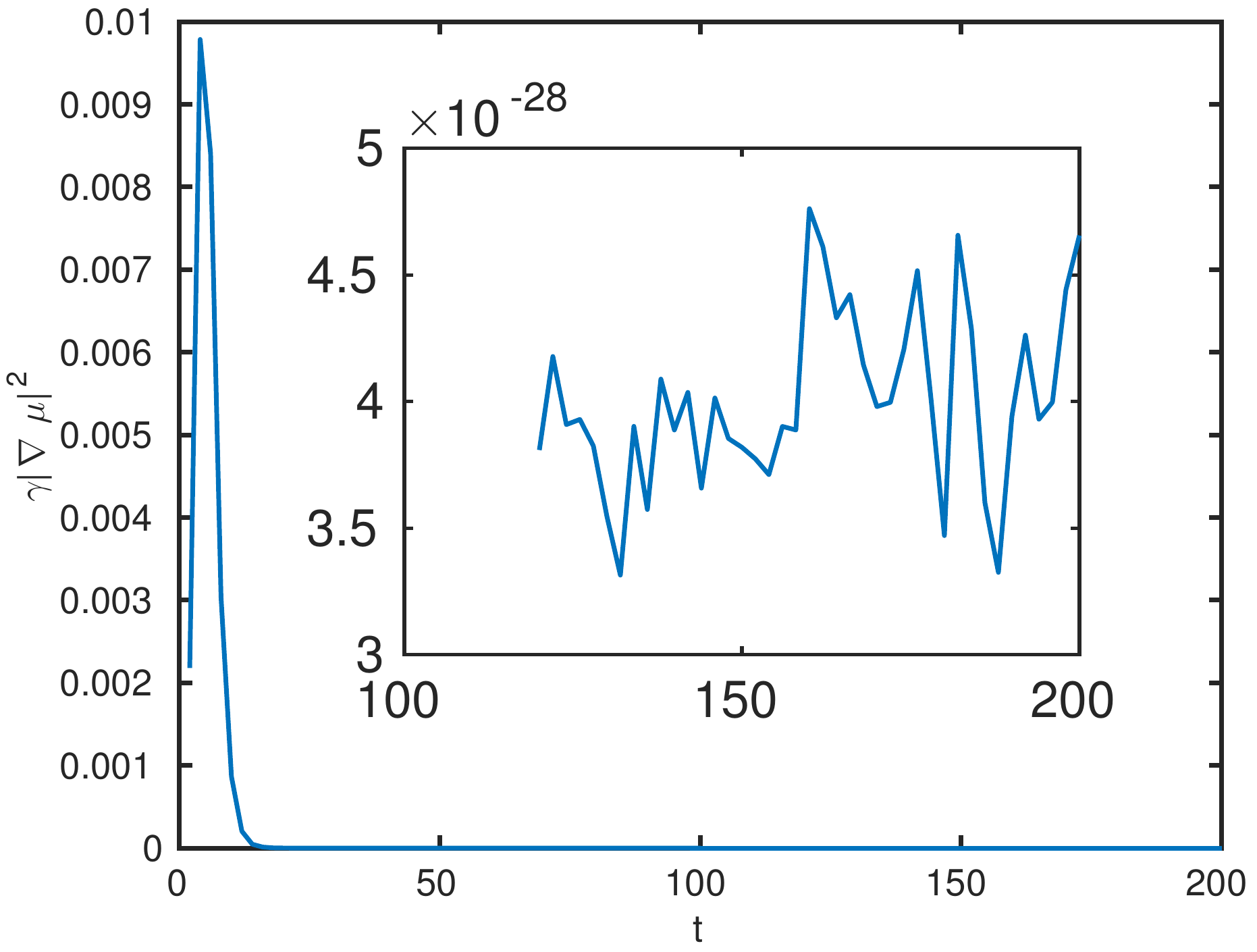}
\includegraphics[width=0.32\linewidth]{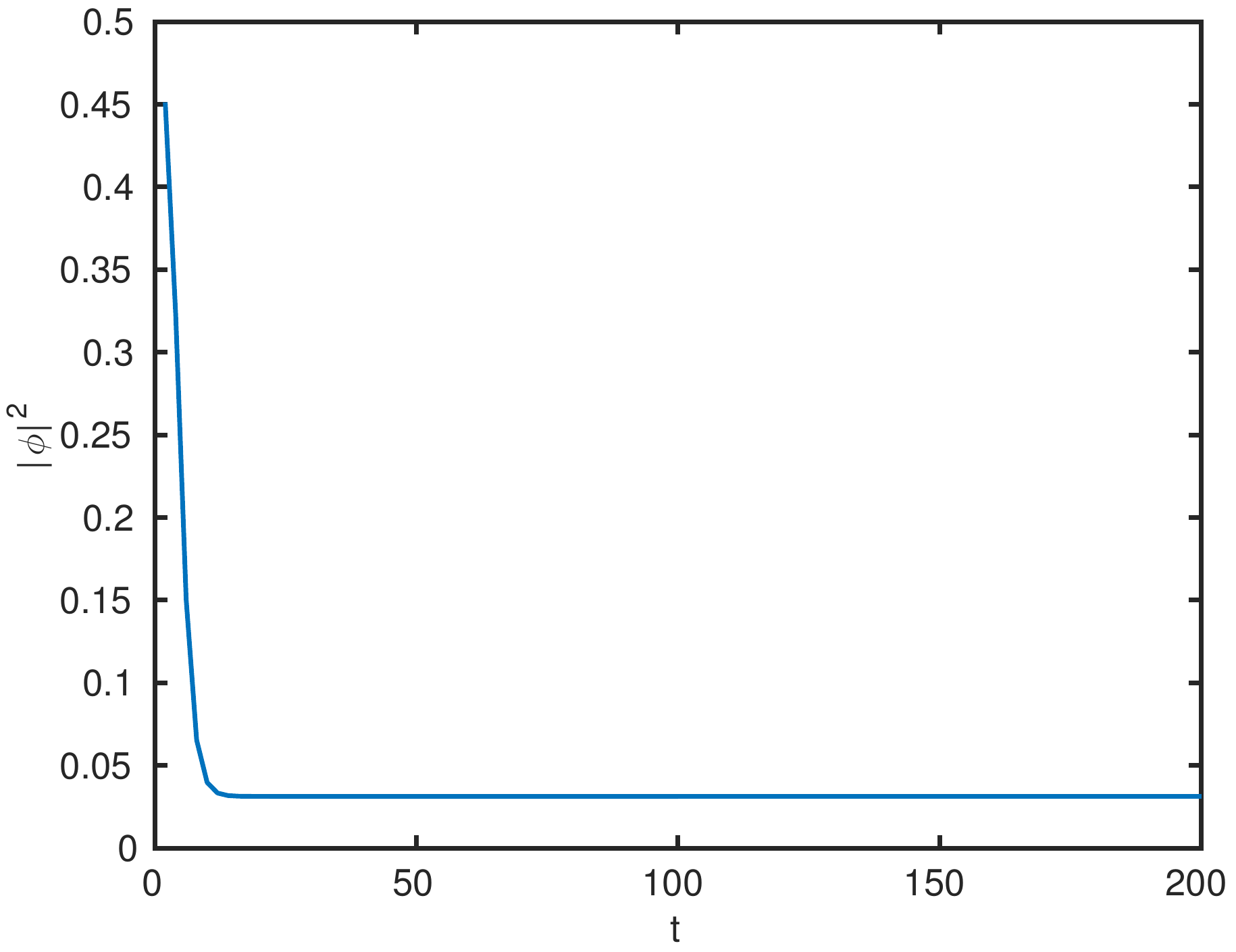}
\includegraphics[width=0.32\linewidth]{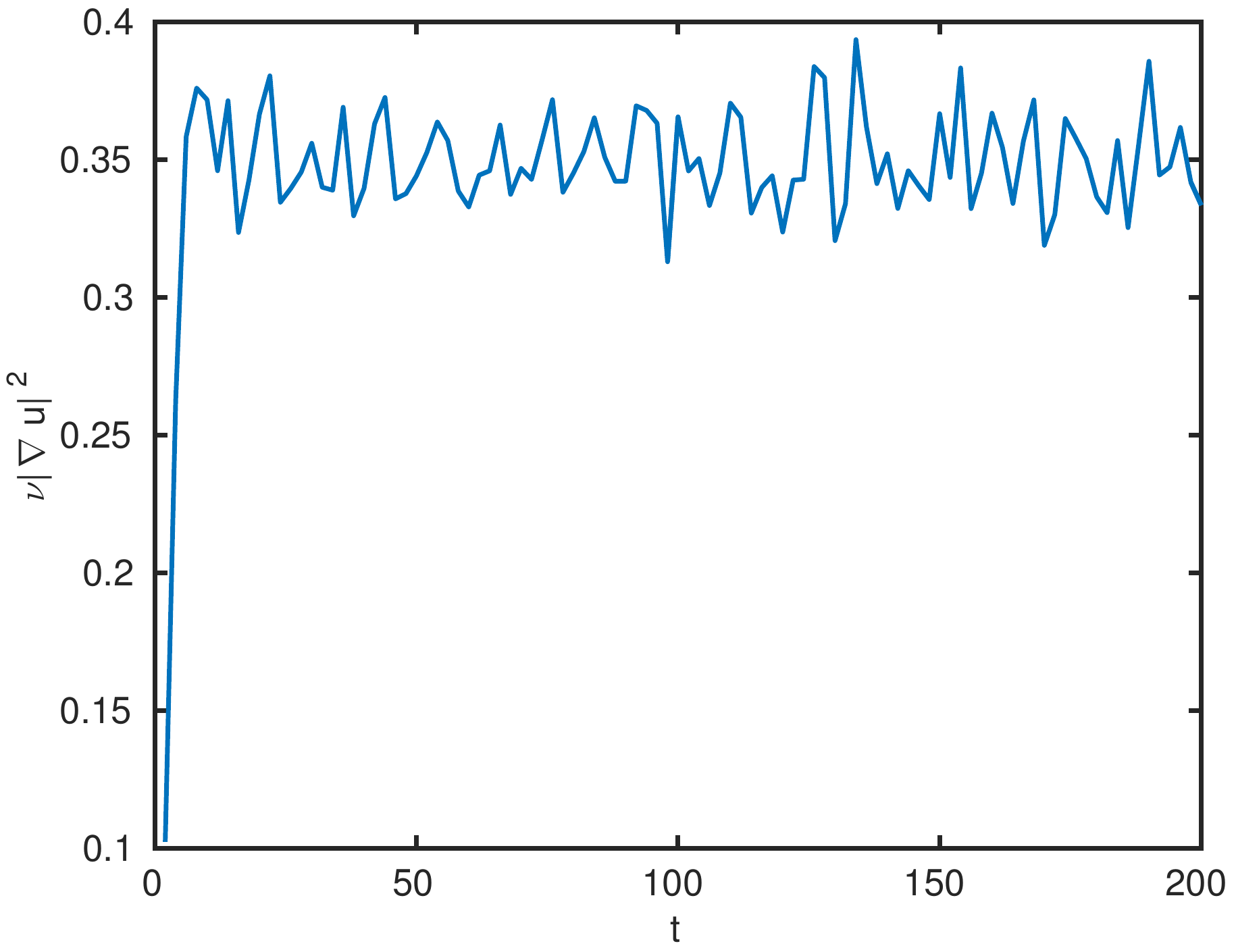}
\caption{Plots versus time $t$ of $\gamma\|\nabla \mu\|_{2}^2$ (left panel), $\|\phi\|_{2}^2$ (middle panel), 
and $\nu\|\nabla\bu\|_{2}^2$ (right panel) for run ${\tt R7}$. The inset in the left panel shows the plot 
of $\gamma\|\nabla\mu\|_{2}^2$ versus time $t$\,; this shows that $\gamma\|\nabla\bu\|_{2}^2$ is vanishingly small 
for this run, in the statistically steady state. The labelling in the figures is $|\cdot|^{2}\equiv 
\|\cdot\|_{2}^{2}$.}\label{fig:R7}
\end{figure}

\smallskip
In the way $\lambda_{\nu,\gamma}$ and $\lambda_{\Lambda,\gamma}$ have been defined, their relative value 
will depend only on $\sqrt\Lambda$ and $\nu$. {\tt R1-R8} in Table~\ref{tableDNS}, remarkably show that, 
for the parameters we have used, 
\bel{order}
\lambda_{\nu,\gamma} \le \lambda_{\Lambda,\gamma}\,.
\ee
Only if $\nu$ = $\gamma$ = $\sqrt{\Lambda}$, do we find $\lambda_{\nu,\gamma} = \lambda_{\Lambda,\gamma}$ 
(see Run {\tt R7}-{\tt R8}). In Figs.~\ref{fig:R1}-\ref{fig:R7}, we have shown the time series 
of $\gamma\ |\nabla \mu|^2$ (left panel), $|\phi|^2$ (middle panel), and  $\nu|\nabla u|^2$ (right panel) 
for run $\tt R1$, $\tt R3$ and $\tt R7$ respectively. For runs $\tt R3, R4, R7, R8$ we have chosen a small 
value of the mixing-energy density, which leads to high mixing and is responsible for the vanishingly small 
values of $\gamma\ |\nabla \mu|^2$ shown in the insets of the left panels of Figs.~\ref{fig:R3} and \ref{fig:R7}.
\par\smallskip
\begin{table*}[h]
\resizebox{1.0\linewidth}{!}
{
\begin{tabular}{|l|l|l|l|l|l|l|l|l|l|l|l|}
\hline
&  $\left<Re_{\lambda}\right>_{t}$ & $\nu$ & $\Lambda$ & $\gamma$ & $\nu\left<\|\nabla\bu\|_{2}^{2}\right>_t$ & 
$\gamma\left<\|\nabla\mu\|_{2}^{2}\right>_t$ & $\lambda_{\nu, \gamma}$ & $\lambda_{\Lambda,\gamma}$\\
\hline
\hline
{\tt R1} &  $27$ & $1.16 \times 10^{-2}$ & $0.107703$ & $1.16 \times 10^{-2}$ & $0.2418$ & $0.1078$ & $1.1462$ & $6.6594$\\
\hline
{\tt R2} &  $27$ & $1.16 \times 10^{-2}$ & $0.107703$ & $6.25 \times 10^{-3}$ & $0.2440$ & $0.1063$ & $1.1456$ & $6.6447$\\
\hline
{\tt R3} &  $30$ & $1.16 \times 10^{-2}$ & $0.0016  $ & $1.16 \times 10^{-2}$ & $0.3496$ & $0$ & $1.1462$ & $2.1285$\\
\hline
{\tt R4} &  $30$ & $1.16 \times 10^{-2}$ & $0.0016  $ & $6.25 \times 10^{-3}$ & $0.3493$ & $0$ & $1.1458$ & $2.1286$\\
\hline
{\tt R5} &  $48$ & $5.00 \times 10^{-3}$  & $0.0707$ & $5.00 \times 10^{-3}$ & $0.1910$ & $0.1589$ & $0.6096$ & $5.1519$\\
\hline
{\tt R6} &  $47$& $5.00 \times 10^{-3}$ & $0.107703$ & $5.00 \times 10^{-3}$ & $0.2184$ & $0.1317$ & $0.6095$ & $5.5441$\\
\hline
{\tt R7} &  $21$& $2.00 \times 10^{-2}$ & $4.00 \times 10^{-4}$ & $2.00 \times 10^{-2}$ & $0.3500$ & $0$ & $1.7249$ & $1.7249$\\
\hline
{\tt R8} &  $13$& $4.00 \times 10^{-2}$ & $1.60 \times 10^{-3}$ & $4.00 \times 10^{-2}$ & $0.3505$ & $0$ & $2.8987$ & $2.8987$\\
\hline
\end{tabular}}
\caption{\label{tableDNS} \small The entries in the table are values of the parameters $Re_{\lambda}$, $\nu$, 
$\Lambda$, $\gamma$, $\nu\left<\|\nabla\bu\|_{2}^2 \right>_{t}$, $\gamma\left<\|\nabla\mu\|_{2}^2 \right>_{t}$, 
$\lambda_{\nu,\gamma}$, and $\lambda_{\Lambda,\gamma}$ for our DNS runs {\tt R1-R8}. The number of 
collocation points is kept fixed at $N=128$ in each direction (so the total number of collocation 
points is $128^3$). The forcing wave numbers are fixed at $k_{f}=1 \& 2$, $\nu$ is the kinematic 
viscosity, the Cahn number $Ch=\xi/L$, where $\xi$ is the interface width, is kept fixed at $Ch=0.01$ 
(with $L=2\pi$), and $Re_{\lambda}$ is the Taylor-microscale Reynolds number. In all cases $\left<\cdot\right>_t$ 
denotes the average over time in the statistically steady state.}
\end{table*}

\subsection{Dimensionless forms of the CHNS equations}\label{dimCHNS}

Let us transform the CHNS equations into dimensionless (primed) coordinates beginning with a characteristic 
velocity $U$ and using the layer thickness $\xi$ as the characteristic length (see Table 2) 
\bel{dim1}
t' = U\xi^{-1}t\,,\qquad x' = \xi^{-1} x\,,\qquad \bu' = U^{-1}\bu\,,\qquad \phi' = \phi\,.
\ee
Moreover, let
\bel{dim2}
\mu' = - \Delta' \phi' + \phi'^{3} - \phi'\,,\qquad\mbox{with}\qquad \mu = \Lambda\xi^{-2} \mu'\,,
\ee
and let the dimensionless pressure $p'$ be defined as $p' = (\tau\xi^{-1})^{2}p$. Then, in dimensionless form, 
the CHNS equations become (with $\mbox{div}'\bu' = 0$)
\beq{dim3}
\left(\partial_{t'} + \bu'\cdot\nabla'\right)\phi' &=& S_{1}^{-1}\Delta'\mu'\,,\\
\left(\partial_{t'} + \bu'\cdot\nabla'\right)\bu' &=&  S_{2}^{-1}\Delta'\bu' - S_{3}^{-1}\phi'\nabla'\mu' 
- \nabla' p' + \bdf'\,,
\eeq
where $S_{1}^{-1}$, $S_{2}^{-1}$ and $S_{3}^{-1}$ are dimensionless parameters\,:
\bel{dim4}
S_{1}^{-1} = \frac{\Lambda\gamma}{U\xi^3}\,,
\qquad
S_{2}^{-1} = \frac{\nu}{U\xi}\,,
\qquad
S_{3}^{-1} = \frac{\Lambda}{U^{2}\xi^2}\,.
\ee
The cubic domain is $[0,\,Ch^{-1}]^{3}$ where $Ch = \xi/L$ is the Cahn number which represents the interface 
thickness normalized with the characteristic length scale (here the characteristic length scale is the 
box-size $L=2\pi$).  In the CHNS literature, the P\'eclet number $Pe$ is also used, which is the ratio 
of the convective and diffusive time scales. It is also the product of the Reynolds number and the Schmidt 
number. For  characteristic length and velocity scales $L$ and $U$\,, $Pe =  LU/D$\,, where $D = \gamma\Lambda/
\xi^2$ is the diffusivity. If the CHNS equation is non-dimensionalized by using $\xi$ as the characteristic 
length scale, then $Pe = U \xi^{3}/\Lambda\gamma = S_1$.  
\par\smallskip
We can also write $S_3 = Ca Re_\xi$, where $Ca =  \rho \nu U/ \sigma$\, is the capillary 
number and $Re_\xi = U \xi/\nu$\, is the Reynolds number at the length scale $\xi$;
and $\sigma  = 2√2 \Lambda/3 \xi$ $\Rightarrow$ $S_3 \sim \rho U^2 \xi^2 /\Lambda$.

\subsection{Proof of Theorem \ref{thmCHNS} for the full set of parameters}\label{CHNSBKM}

The number of parameters $S_{i}^{-1}$ ($i=1,2,3$) in (\ref{dim3}) significantly lengthens and 
complicates the proof of Theorem \ref{thmCHNS} so in \cite{Gibbon2017PRE} this was performed 
for unit parameters\,: this is repeated in \ref{appB} for completeness. In this section we 
show that there is a way of adapting the unit-parameter proof to the full set of parameter 
values. Let us return to the energy 
\beq{dim7}
E(t) &=& \IV \left\{\shalf\Lambda|\nabla\phi|^{2} + \frac{\Lambda}{4\xi^{2}} (\phi^{2} - 1)^{2} 
+ \shalf|\bu|^{2}\right\}dV\nonumber\\
&=&  \xi\Lambda\IV \left\{\shalf|\nabla'\phi'|^{2} + \quart (\phi'^{2} - 1)^{2} 
+ \shalf S_{3}|\bu'|^{2}\right\}dV'\nonumber\\
&\equiv& \xi\Lambda E'(t)\,,
\eeq
where, based on the definition of $E'(t)$ in (\ref{dim7}), we define its $L^{\infty}$-equivalent
\bel{dim8}
E'_{\infty} = \shalf\|\nabla'\phi'\|_{\infty}^{2} + \quart (\|\phi'\|_{\infty}^{2} - 1)^{2} 
+ \shalf S_{3}\|\bu'\|_{\infty}^{2}\,.
\ee
and thus $E'_{\infty} = \Lambda^{-1}\xi^{2}E_{\infty}$ with $E_{\infty}$ defined in (\ref{Einfintro}).  
Indeed, the calculation leading to (\ref{Ecal3}) can be repeated using the dimensionless form $E'$ 
above. In the following we develop a strategy based upon $E'_{\infty}$ defined in (\ref{dim8}). 
In \cite{Gibbon2017PRE} a BKM-type theorem was proved with the various parameters set to unity for 
convenience, which makes $S_{1} = S_{2} = S_{3}$. This proof in \cite{Gibbon2017PRE} can be used 
to prove the theorem for the full parameter set by using a device. Firstly, it is easy to prove 
that 
\bel{dim9}
\min\left\{1,S_{3}^{-1}\right\} E'_{\infty} \leq 
E'_{unit} \leq \max\left\{1,S_{3}^{-1}\right\} E'_{\infty} \,.
\ee
where, with unit variables, 
\bel{Eprimeunit}
E'_{unit} = \shalf\|\nabla'\phi'\|_{\infty}^{2} + \quart (\|\phi'\|_{\infty}^{2} - 1)^{2} 
+ \shalf \|\bu'\|_{\infty}^{2}\,.
\ee
The direction of the inequalities in the proof allow us to prove the theorem in the primed variables 
with unit parameter values (i.e. $E'_{unit}$), and then, using (\ref{dim9}), replace 
$E'_{unit}$ with  $E'_{\infty}$, which translates back to $E_{\infty}$ in dimensional variables.
Thus we have proved the theorem for all positive values of the parameters in $E_{\infty}$. 


\section{Conclusion}

The regularity problem for the $3D$ CHNS equations is a hard problem\,: it compounds the formidable difficulties 
found when addressing the same issue in its two constituent parts, namely the $3D$ Navier-Stokes and $3D$ CHNS equations respectively. While there are also clear parallels with the Navier-Stokes problem, which suggests a Leray-type 
approach to weak solutions might be fruitful, various difficulties stand in the way. For instance, in (\ref{Ecal3}) 
the usual bound on the time average of $\nu\IV|\nabla\bu|^{2}\,dV$ in the energy dissipation rate is the root of all the 
results for the Navier-Stokes part, but we also have the Cahn-Hilliard contribution of $\gamma\IV|\nabla\mu|^{2}\,dV$ 
on the right hand side. Finding estimates in terms of this is a difficult task and one that currently lies out of reach.

For the present we have to be content with a Beale-Kato-Majda-type of result as displayed in Theorem \ref{thmCHNS}. 
We could now call this a class of theorems as there are now three of its type\,: (i) the original for the $3D$ Euler 
equations\,; (ii) our Theorem \ref{thmCHNS} for the CHNS equations, and (iii) a theorem for the stochastic $3D$ Euler 
equations of Crisan, Flandoli and Holm \cite{CFH2017}. The $128^3$ simulations display no evidence of singular 
behaviour although, computationally, this is a very demanding problem and requires further investigation. 
\par\smallskip
The structure of the energy dissipation in (\ref{Ecal3}) has allowed us to introduce an $Re^3$ energy bound 
and thus a $Re^{3/4}$ upper bound on $(L\lambda_{\nu,\gamma})^{-1}$, this inverse length scale being defined 
in (\ref{Kol1}). One further interesting result arising in the simulations is the fact that for both $\nu > 
\sqrt{\Lambda}$ and $\nu < \sqrt{\Lambda}$ we see an ordering in these two length scales such that 
$\lambda_{\nu,\gamma} < \lambda_{\Lambda,\gamma}$, a result for which we see no evidence analytically. This, 
again, requires further investigation.


\section*{Acknowledgements}
RP and NP thank the Department of Science and Technology, Council for Scientific and Industrial Research, 
the University Grants Commission (India), and the Indo-French Centre for Applied Mathematics (IFCAM) for
support, and SERC (IISc) for computational resources. The authors are grateful to the referees for some 
suggested revisions.


\appendix
\section{The Doering-Foias relation between $Gr$ and $Re$}\label{appA}

Doering and Foias \cite{DF2002} split the forcing function $\bdf(\bx)$ into its magnitude $F$ and 
its ``shape'' $\bphi$ such that
\bel{p1}
\bdf(\bx) = F\bphi(\ell^{-1}\bx) ,
\ee
where $\ell$ is the longest length scale in the force but here it is taken to be $\ell = L$ for convenience. 
On the unit torus $\mathbb{I}_{d}$, $\bphi$ is a mean-zero, divergence-free vector field with the chosen 
normalization property 
\bel{p1a}
\int_{\mathbb{I}_{d}} \left|\nabla^{-1}_{y}\bphi\right|^{2}\,d^{d}y = 1\,.
\ee
$L^{2}$-norms of $\bdf$ on $\mathbb{I}^{d}$ are 
\bel{p2}
\|\nabla^{N}\bdf\|_{2}^{2} = C_{N}\ell^{-2N} L^{d} F^{2} ,
\ee
where the coefficients $C_{N}$ refer to the shape of the force but not its magnitude
\bel{p3}
C_{M} = \sum_{n} \left| 2\pi n\right|^{2N} |\hat{\bphi}_{n}|^{2}\,.
\ee
Doering and Foias~\cite{DF2002} showed that various bounds exist such as (among others)
\bel{p4}
\|\nabla\Delta^{-M}\bdf\|_{\infty} = D_{M}F \ell^{2M-1}\,.
\ee
The energy-dissipation rate $\epsilon$ is  
\bel{p5}
\epsilon = \left<\nu L^{-d}\int_{\mathbb{I}_{d}} \ |\nabla\bu|^{2}\,dV\right> = \nu L^{-d}\left<H_{1}\right>\,.
\ee
In terms of $F$ the Grashof number in (\ref{GrRedef}) becomes
\bel{Grdef}
Gr = F\ell^{3}/\nu^{2}
\ee
and the Taylor micro-scale $\lambda_{T}$ is related to $U$ via $\lambda_{T} = \sqrt{\nu U^{2}/\epsilon}$\,,
which is consistent with the definition $\lambda_{T}^{-2} = \left<H_{1}\right>/\left<H_{0}\right>$.
\par\bigskip\noindent
Following the procedure in~\cite{DGbook1995} (page 296 equation (2.9)) we multiply the Navier-Stokes equations 
by $(-\Delta^{-M})\bdf$ to obtain
\beq{a1}
\frac{d~}{dt}\int_{\mathbb{I}_{d}} \bu\cdot[(-\Delta^{-M})\bdf]\,dV 
&=& \nu\int_{\mathbb{I}_{d}} \Delta\bu\cdot[(-\Delta^{-M})\bdf]\nonumber\\
&-& \int_{\mathbb{I}_{d}} \nabla^{-M}\bdf\cdot \nabla^{-M}\bdf \,dV\nonumber\\
&-& \int_{\mathbb{I}_{d}} \bu\cdot\nabla\bu \cdot[(-\Delta^{-M})\bdf]\,dV\nonumber\\
&+& \IV\phi(\Delta^{-M}\bdf)\cdot\nabla\mu\,dV\,,
\eeq
where the pressure term vanishes in the usual way. Now integrate all the terms by parts, and take the time average
\beq{a2}
\left<\int_{\mathbb{I}_{d}} \nabla^{-M}\bdf\cdot \nabla^{-M}\bdf \,dV\right>_{T} 
&\leq& 
\nu\left<\int_{\mathbb{I}_{d}}\bu\cdot[(-\Delta^{-M+1})\bdf]\,dV\right>_{T}\nonumber\\
-\left<\int_{\mathbb{I}_{d}}\bu\cdot[\nabla[(-\Delta^{-M})]\bdf]\cdot\bu\,dV\right>_{T}
&+& \left< \int_{\mathbb{I}_{d}}\phi \Delta^{-M}\bdf\cdot\nabla\mu\right>_{T} .
\eeq
An extra term $\IV\phi \Delta^{-M}\bdf\cdot\nabla\mu\,dV$ derives from the $-\phi\nabla\mu$-term 
in (\ref{chns1}). However, all its contributions are zero except one, given the definition of $\mu$. 
(\ref{a2}) becomes 
\bel{a4}
c_{0}F^{2}\ell^{2M}\leq c_{1}\nu F\ell^{2M-2}U + c_{2}\ell^{2M-1}F U^{2} + 
c_{3}\Lambda^{-1}\ell^{2M-1}F \left<E\right>_{T}\,,
\ee
where the $U^2$-term contains the contributions from both nonlinear terms and the constants (not 
explicitly given) contain the shape of the body forcing. Using (\ref{Grdef}), as $Gr \to\infty$, 
(\ref{a4}) becomes 
\bel{a5}
Gr \leq c_{4}\left(Re + Re^{2}\right) + O\left(\left<E\right>_{T}\right)\,.
\ee


\section{Proof of Theorem \ref{thmCHNS} with unit parameters}\label{appB}

In the following the parameters in the dimensionless system $S_{n}$ are set to unity and primes have been 
removed\footnote{This proof has been published in \cite{Gibbon2017PRE} but is included here for completeness.}. 
The domain is now the cube $[0,\,Ch^{-1}]^{3}$. Then, we recall the definitions of $H_{n}$ in equation 
(\ref{Hndef}) and $P_{n}$ in equation (\ref{Pndef}) and proceed in 3 steps. 
\par\medskip\noindent
\textbf{Step 1\,:} We begin with the time evolution of $P_{n}$\, (the dot above $P_n$ denotes a time 
derivative),:
\bel{pn1}
\shalf\dot{P}_{n} =  - P_{n+2} + P_{n+1} + \IV (\nabla^{n}\phi)\nabla^{n}\Delta(\phi^{3})\,dV
- \IV (\nabla^{n}\phi )\nabla^{n}(\bu\cdot\nabla\phi)\,dV\,;
\ee
and then we estimate the third term on the right as
\bel{pn2}
\left|\IV (\nabla^{n}\phi)\nabla^{n}\Delta(\phi^{3})\,dV\right| \leq
\|\nabla^{n}\phi\|_{2}\sum_{i,j=0}^{n+2}C^{n+2}_{i,j}\|\nabla^{i}\phi\|_{p}
|\nabla^{j}\phi\|_{q}\|\nabla^{n+2-i-j}\phi\|_{r},
\ee
where $1/p + 1/q + 1/r = 1/2$. Now we use a sequence of Gagliardo-Nirenberg inequalities 
\beq{pn3}
\|\nabla^{i}\phi\|_{p} &\leq& c_{n,i} \|\nabla^{n+2}\phi\|_{2}^{a_{1}}\|\phi\|_{\infty}^{1-a_{1}},\non\\
\|\nabla^{j}\phi\|_{q} &\leq& c_{n,j} \|\nabla^{n+2}\phi\|_{2}^{a_{2}}\|\phi\|_{\infty}^{1-a_{2}},\\
\|\nabla^{n}\phi\|_{r} &\leq& c_{n,i,j} \|\nabla^{n+2-i-j}\phi\|_{2}^{a_{3}}\|\phi\|_{\infty}^{1-a_{3}}\,,\non
\eeq
where, in $d$ dimensions,
\beq{pn4}
\frac{1}{p} &=& \frac{i}{d} + a_{1}\left(\frac{1}{2} - \frac{n+2}{d}\right),\non\\
\frac{1}{q} &=& \frac{j}{d} + a_{2}\left(\frac{1}{2} - \frac{n+2}{d}\right),\\
\frac{1}{r} &=& \frac{n+2-i-j}{d} + a_{3}\left(\frac{1}{2} - \frac{n+2}{d}\right)\,.\non
\eeq
By summing these and using $1/p + 1/q + 1/r = 1/2$, it is seen that $a_{1}+a_{2}+a_{3}=1$. Thus, we have 
\bel{pn5}
\left|\IV (\nabla^{n}\phi)\nabla^{n+2}(\phi^{3})\,dV\right| \leq
c_{n}\|\nabla^{n}\phi\|_{2}\|\nabla^{n+2}\phi\|_{2}\|\phi\|_{\infty}^{2}
\leq \shalf P_{n+2} + c_{n} P_{n}\|\phi\|_{\infty}^{4}\,,
\ee
and so Eq.~\eqref{pn1} becomes (here and henceforth coefficients such as $c_n$ are multiplicative constants),
\bel{pn6}
\shalf\dot{P}_{n}  =  -\shalf P_{n+2} + P_{n+1} + c_{n}\|\phi\|_{\infty}^{4}P_{n} + 
\left|\IV (\nabla^{n}\phi)\nabla^{n}(\bu\cdot\nabla\phi)\,dV\right|\,.
\ee
Estimating the last term in Eq.~\eqref{pn6} we have
\beq{pn7}
\left|\IV (\nabla^{n}\phi) \nabla^{n}(\bu\cdot\nabla\phi)\,dV\right| &=& 
\left|-\IV (\nabla^{n+1}\phi) \nabla^{n-1}(\bu\cdot\nabla\phi)\,dV\right|\\
 &\leq& \|\nabla^{n+1}\phi\|_{2} \sum_{i=0}^{n-1}C^{n}_{i}\|\nabla^{i}\bu\|_{p}
\|\nabla^{n-1-i}(\nabla\phi)\|_{q}\,,\non
\eeq
where $1/p + 1/q = 1/2$. Now we use two Gagliardo-Nirenberg inequalities in $d$ dimensions to obtain
\beq{pn8}
\|\nabla^{i}\bu\|_{p} &\leq& c\,\|\nabla^{n-1}\bu\|_{2}^{a} \|\bu\|_{\infty}^{1-a},\\
\|\nabla^{n-1-i}(\nabla\phi)\|_{q} & \leq & c\,\|\nabla^{n-1}(\nabla\phi)\|_{2}^{b} \|\nabla\phi\|_{\infty}^{1-b}.
\label{pn8b}
\eeq
Equations~\eqref{pn8} and \eqref{pn8b} follow from
\beq{lb1}
\frac{1}{p} & = & \frac{i}{d} + a\left(\frac{1}{2} - \frac{n-1}{d}\right),\\
\frac{1}{q} & =  & \frac{n-1-i}{d} + b\left(\frac{1}{2} - \frac{n-1}{d}\right).
\eeq
Because $1/p + 1/q = 1/2$ then $a+b=1$. Thus the second term in Eq.~\eqref{pn1} turns into\footnote{Inequalities 
(\ref{pn8}) and (\ref{pn9}) are the origin of the $\|\bu\|_{\infty}^{2}$-term in $E_{\infty}$.} 
\beq{pn9}
\left|\IV (\nabla^{n}\phi)\nabla^{n}(\bu\cdot\nabla\phi)\,dV\right| &\leq&
c_{n}P_{n+1}^{1/2} H_{n-1}^{a/2}P_{n}^{b/2}\|\bu\|_{\infty}^{1-a}\|\nabla\phi\|_{\infty}^{1-b}\\
&\leq& P_{n+1}^{1/2} \left[c_{n}H_{n-1}\|\nabla\phi\|_{\infty}^{2}\right]^{a/2}\left[P_{n}\|\bu\|_{\infty}^{2}\right]^{b/2}\non\\
&\leq& \shalf P_{n+1} + \shalf ac_{n}H_{n-1}\|\nabla\phi\|_{\infty}^{2}+ \shalf b P_{n}\|\bu\|_{\infty}^{2}\non\,,
\eeq
and Eq.~\eqref{pn6} becomes
\bel{pn10}
\shalf\dot{P}_{n}  =  - \shalf P_{n+2} + \threehalves P_{n+1} 
+ c_{n,1}\left(\shalf\|\phi\|_{\infty}^{4} + \|\bu\|_{\infty}^{2}\right)P_{n} 
+ c_{n,2}H_{n-1}\|\nabla\phi\|_{\infty}^{2}\,.
\ee
\par\medskip\noindent
\textbf{Step 2\,:} Now we look at $H_{n}$ defined in Eq.~\eqref{Hndef} using Eq.~\eqref{nsedef} with 
$\bdf = - \hat{z}\phi$. The easiest way is to use the 3D NS equation in the vorticity form as in 
Doering and Gibbon \cite{DGbook1995} where gradient terms have been absorbed into the pressure term, 
which disappears under the curl-operation
\bel{h13d}
\left(\partial_{t}+\bu\cdot\nabla\right)\bom = \Delta\bom + \bom\cdot\nabla\bu + \nabla\phi\times\nabla\Delta\phi 
- \nabla^{\perp}\phi\,.
\ee
Therefore, following the methods used in \cite{DGbook1995}, we find
\beq{hn1}
\shalf\dot{H}_{n} &\leq& -\shalf H_{n+1} + c_{n}\|\bu\|_{\infty}^{2}H_{n} + 
\left|\IV(\nabla^{n-1}\bom)\left[\nabla^{n-1}\left(\nabla\phi\times\Delta\nabla\phi\right)\right]\,dV\right|\non\\
&+& \left|\IV (\nabla^{n-1}\bom)\left[\nabla^{n-1}\nabla^{\perp}\phi\right]\,dV\right|\,.
\eeq
Beginning with the third term on the right-hand side of Eq.~\eqref{hn1}, we obtain 
\bel{hn2}
\left|\IV (\nabla^{n-1}\bom)\nabla^{n-1}\left(\nabla\phi\times\Delta\nabla\phi\right)\,dV\right|
\leq \|\nabla^{n-1}\bom\|_{2} \sum_{i=0}^{n-1}C^{n}_{i}\|\nabla^{i}(\nabla\phi)\|_{r}\|\nabla^{n+1-i}
(\nabla\phi)\|_{s}\,.
\ee
Then, by using a Gagliardo-Nirenberg inequality,
\beq{hn3}
\|\nabla^{i}(\nabla\phi)\|_{r} &\leq& c\,\|\nabla^{n+1}(\nabla\phi)\|_{2}^{a} \|\nabla\phi\|_{\infty}^{1-a},\\
\|\nabla^{n+1-i}(\nabla\phi)\|_{s}
&\leq& c\,\|\nabla^{n+1}(\nabla\phi)\|_{2}^{b} \|\nabla\phi\|_{\infty}^{1-b}\,,
\eeq
where $1/r + 1/s = 1/2$ and where
\beq{eq2}
\frac{1}{r} & = & \frac{i}{d} + a\left(\frac{1}{2} - \frac{n+1}{d}\right)\\
\frac{1}{s} & = & \frac{n+1-i}{d} + b\left(\frac{1}{2} - \frac{n+1}{d}\right)\,,
\eeq
we find that $a+b=1$. This yields
\beq{hn4}
\left|\IV (\nabla^{n-1}\bom)\nabla^{n-1}\left(\nabla\phi\times\Delta\nabla\phi\right)dV\right|
&\leq& c_{n}H_{n}^{1/2}P_{n+2}^{1/2}\|\nabla\phi\|_{\infty}\non\\
&\leq& P_{n+2} + \quart c_{n}H_{n}\|\nabla\phi\|_{\infty}^{2}\,.
\eeq
The last term on the right-hand side of Eq.~\eqref{hn1} is easily handled. Altogether we find 
\beq{hn5}
\shalf\dot{H}_{n} & \leq & -\shalf H_{n+1} + P_{n+2} + c_{n,3}\left(\|\bu\|_{\infty}^{2} 
+ \|\nabla\phi\|_{\infty}^{2}\right)H_{n} + \shalf H_{n} + \shalf P_{n}\,.
\eeq
\par\medskip\noindent
\textbf{Step 3\,:} Finally, by noting that $X_{n} =  P_{n+1} + H_{n}$, we use 
Eq.~\eqref{pn5} with $n \to n+1$ to obtain
\beq{hn6}
\shalf \dot{X}_{n} &\leq& - \shalf P_{n+3} + \threehalves P_{n+2} + 
c_{n,1}\left(\shalf\|\phi\|_{\infty}^{4} + \|\bu\|_{\infty}^{2}\right)P_{n+1} 
+ c_{n,2}H_{n}\|\nabla\phi\|_{\infty}^{2}\non\\
&-& \shalf H_{n+1} + P_{n+2} + c_{n,3}\left(\|\bu\|_{\infty}^{2} + \|\nabla\phi\|_{\infty}^{2}\right)H_{n} 
+ \shalf H_{n} + \shalf P_{n}\non\\
&\leq& - \shalf P_{n+3} - \shalf H_{n+1} + \fivehalves P_{n+2} +
c_{n,4}\left(\shalf\|\phi\|_{\infty}^{4} + \|\bu\|_{\infty}^{2} + 
\|\nabla\phi\|_{\infty}^{2}\right)X_{n} \non\\
&+& \shalf H_{n} + \shalf P_{n}\,.
\eeq
By using $P_{n+2} \leq P_{n+3}^{1/2}P_{n+1}^{1/2} \leq (\varepsilon/2) P_{n+3} + 
(1/2\varepsilon)P_{n+1}$, with $\varepsilon$ chosen as $\varepsilon = \onefifth$, we have 
(with $P_{n} \leq Ch^{-2}P_{n+1}$)
\bel{hn7}
\shalf \dot{X}_{n} \leq - \quart P_{n+3} - \shalf H_{n+1} + 
c_{n,4}\max(1,\,Ch^{-2})\left(\|\nabla\phi\|_{\infty}^{2} + \shalf\|\phi\|_{\infty}^{4} + \|\bu\|_{\infty}^{2} 
+ \shalf\right)X_{n}\,.
\ee
We note that $\phi$ is a mean-zero function on our periodic domain $[0\,,Ch^{-1}]^{3}$, so $\|\phi\|_{\infty}
\leq Ch^{-1}\|\nabla\phi\|_{\infty}$. Then we can write
\beq{hn8}
&c_{n,5}&\left(\|\nabla\phi\|_{\infty}^{2} + \shalf\|\phi\|_{\infty}^{4} + \|\bu\|_{\infty}^{2} + \shalf\right) 
=  \non\\ 
&c_{n,5}&\left(\|\nabla\phi\|_{\infty}^{2} + \shalf\left(\|\phi\|_{\infty}^{2}-1\right)^{2}+\|\bu\|_{\infty}^{2} 
+ \|\phi\|_{\infty}^{2}\right)\non\\
&\leq& 2c_{n,5}\left(\|\nabla\phi\|_{\infty}^{2} + \shalf\left(\|\phi\|_{\infty}^{2}-1\right)^{2} 
+ \|\bu\|_{\infty}^{2}\right)\,.
\eeq
By dropping the negative terms, Eq.~\eqref{hn7} turns into 
\bel{hn8}
\quart \dot{X}_{n} \leq c_{n,5}E'_{unit} X_{n}\,,
\ee
where $E'_{unit}$ is defined in Eq.~\eqref{dim9} which has unit parameters. This can then be 
replaced by $E'_{\infty}$ using the same inequality. By integrating over $[0,\,T^{*}]$, we obtain
\bel{Xnexp}
X_{n}(T^{*}) \leq c_{n,6}X_{n}(0)\exp \int_{0}^{T^{*}}E'_{\infty}(\tau)\,d\tau\,.
\ee
Clearly, having $\int_{0}^{T^{*}}E'_{\infty} < \infty$ contradicts the statement in the Theorem that 
solutions first lose regularity at $T^{*}$. Translating back to dimensional variables we have the result. 
\hfill$\blacksquare$


\end{document}